\documentclass[11pt,a4paper]{article}
\pdfoutput=1
\usepackage{jheppub}
\usepackage{bm}

\newcommand{\z}{z}
\DeclareMathOperator{\tr}{tr}

\title{Differential equations on unitarity cut surfaces}
\author{Mao Zeng}
\affiliation{Bhaumik Institute for Theoretical Physics,\\
Department of Physics and Astronomy, University of California,\\
430 Portola Plaza, Los Angeles, USA}
\emailAdd{zengmao@physics.ucla.edu}
\keywords{Perturbative QCD, Scattering Amplitudes}
\abstract{We reformulate differential equations (DEs) for Feynman integrals to avoid doubled propagators in intermediate steps. External momentum derivatives are dressed with loop momentum derivatives to form tangent vectors to unitarity cut surfaces, in a way inspired by unitarity-compatible IBP reduction. For the one-loop box, our method directly produces the final DEs without any integration-by-parts reduction. We further illustrate the method by deriving maximal-cut level differential equations for two-loop nonplanar five-point integrals, whose exact expressions are yet unknown. We speed up the computation using finite field techniques and rational function reconstruction.}
\preprint{UCLA-17-TEP-102}
\arxivnumber{1702.02355}
\begin{document}
\maketitle

\section{Introduction}
\label{sec:intro}
The generalized unitarity method \cite{Bern:1994cg, Bern:1995db, Bern:1996ja, Bern:1997sc, Britto:2004nc, Ossola:2006us, Forde:2007mi, Ellis:2011cr, Ita:2011hi} has been very successfully applied to constructing loop integrands in quantum field theories, while applications to loop integration remain frontiers to explore. Integration-by-parts (IBP) reduction \cite{Chetyrkin:1981qh, Laporta:2001dd, Laporta:1996mq, Smirnov:2014hma, vonManteuffel:2012np} was recently reformulated \cite{Gluza:2010ws, Schabinger:2011dz, Chen:2015lyz, Sogaard:2014jla, Georgoudis:2015hca, Ita:2015tya, Larsen:2015ped, Georgoudis:2016wff, Ita:2016oar, Zhang:2016kfo} in a unitarity-compatible manner without doubled propagators, relying on special combinations of loop momentum derivatives which form tangent vectors \cite{Ita:2015tya} to unitarity cut surfaces.

After IBP reduction to a basis of master integrals, the master integrals still need to be evaluated, e.g.\ using the method of differential equations (DEs) \cite{Kotikov:1990kg, Bern:1993kr, Remiddi:1997ny, Gehrmann:1999as, Argeri:2007up, Henn:2013nsa}. A recent breakthrough was Henn's canonical form of DEs \cite{Henn:2013pwa, Henn:2014qga}, allowing a large class of loop integrals to be expressed in terms of iterated integrals of uniform transcendentality. Various algorithms and software packages \cite{Lee:2014ioa, Henn:2014qga, Tancredi:2015pta, Meyer:2016slj, Meyer:2017joq, Prausa:2017ltv, Gituliar:2017vzm} have appeared to find algebraic transformations of DEs to the canonical form, while a complementary approach is finding master integrals in the $d\log$ form with unit leading singularities \cite{ArkaniHamed:2010gh, Henn:2014qga, Bern:2015ple}.

In constructing DEs, an important intermediate step is IBP reduction, which brings the RHS of the DEs into a linear combination of master integrals. External momentum derivatives increase the power of propagator denominators, so unitarity-compatible IBP reduction is not directly applicable. To solve this problem, one approach is to decrease the power of propagator denominators using dimension shifting \cite{Georgoudis:2016wff}.
However, we propose an alternative approach that completely avoids doubled propagators, even in intermediate steps. We promote unitarity cut surfaces to be objects embedded in the space of not only loop momenta, but also external momenta. By combining external and loop momentum derivatives to form tangent vectors to unitarity cut surfaces, doubled propagators cancel out, in direct analogy with unitarity-compatible IBP reduction. 

It has been proposed that the maximal cut can provide valuable information about differential equations and the function space of the integrals \cite{CaronHuot:2012ab, Primo:2016ebd, Frellesvig:2017aai}. The latter two references rely on consistent definitions of unitarity cuts in the presence doubled propagators (see also \cite{Lee:2012te, Sogaard:2014ila}). This issue is bypassed in our approach by construction.

Section \ref{sec:basic} sketches the basics of our formalism. Section \ref{sec:inverse} uses inverse propagator coordinates, also known as the Baikov representation in the $d$-dimensional case, to present the detailed formalism through the one-loop box example. Section \ref{sec:pentabox} applies the formalism to the nonplanar pentabox at the maximal cut level, and finds a system proportional to the dimensional regularization parameter $\epsilon$, for tensor integrals with unit leading singularities. Section \ref{sec:comp} discusses the use of finite field techniques and rational function construction in speeding up the nonplanar pentabox computation. Some concluding remarks are given in Section \ref{sec:conclusions}.

\section{Basic formalism}
\label{sec:basic}
\subsection{Avoiding doubled propagators}
Generalized unitarity cuts replace propagators by delta functions. However, when the propagator is doubled (i.e.\ squared), it is no longer straightforward to impose unitarity cuts \cite{Primo:2016ebd, Frellesvig:2017aai, Lee:2012te, Sogaard:2014ila}. Therefore, a unitarity-compatible approach to integration-by-parts reduction uses special IBP relations that do not involve doubled propagators \cite{Gluza:2010ws, Schabinger:2011dz, Chen:2015lyz, Sogaard:2014jla, Georgoudis:2015hca, Ita:2015tya, Larsen:2015ped, Georgoudis:2016wff, Ita:2016oar, Zhang:2016kfo}. This allows IBP relations to be put on unitarity cuts, which can be exploited to construct multi-loop generalizations of the OPP parameterization \cite{Ossola:2006us} of one-loop integrands, as well as allowing analytic IBP reduction to be achieved by merging results from a spanning set of unitarity cuts \cite{Larsen:2015ped}. 

We will explore a similar unitarity-compatible approach to differential equations, with no Feynman integrals involving doubled propagators appearing on the LHS or RHS of the differential equations, even before IBP reduction is performed to simplify the RHS. The advantage is two-fold. First, the differential equations may be put on unitarity cuts, and therefore can be constructed by merging incomplete results on a spanning set of unitarity cuts. Second, unitarity-compatible IBP reduction can be used to reduce the RHS into the original set of master integrals, since no integrals with doubled propagators are ever generated.

We wish to compute the derivative of a Feynman integral with propagators $1/\z_j$ and a tensor numerator $\mathcal N$,
\begin{equation}
 \sum_i \beta^\mu_i \frac{\partial}{\partial p_i^\mu} \int d^d l \, \frac{\mathcal N}{\prod_j \z_j} \, ,
\label{eq:oneLoopDeri}
\end{equation}
where $\beta^\mu_i$ is a linear combination of external momenta $p_j$,
\begin{equation}
\beta^\mu_i = \beta_{ij}\, p_j^\mu \, .
\end{equation}

We are free to add total divergences to Eq.\ \eqref{eq:oneLoopDeri} without changing its value after integration, obtaining
\begin{align}
&\quad \int d^d l \, \left[ \sum_i \beta^\mu_i \frac{\partial}{\partial p_i^\mu} \frac{\mathcal N}{\prod_j \z_j} + \frac{\partial}{\partial l^\mu} \frac{ \mathcal N \, v^\mu}{\prod_j \z_j} \right] \label{eq:extPlusInt}\\
&= \int d^d l \, \sum_i \left( \beta^\mu_i \frac{\partial}{\partial p_i^\mu} + v^\mu \frac{\partial}{\partial l^\mu}  \right) \frac{\mathcal N}{\prod_j \z_j} + \int d^d l \frac{\mathcal N}{\prod_j \z_j} \, \frac{\partial v^\mu}{\partial l^\mu} \, .
\label{eq:diffPlusDiv}
\end{align}
In the above expressions, $v^\mu = v^\mu (l,p)$ has polynomial dependence on internal and external momenta, with one free Lorentz index. The final expression Eq.\ \eqref{eq:diffPlusDiv} has \emph{no doubled propagators} $\sim 1/\z_j^2$, if the following condition is satisfied,
\begin{equation}
\left( \beta^\mu_i \frac{\partial}{\partial p_i^\mu} + v^\mu \frac{\partial}{\partial l^\mu}  \right) \z_j = f_j\, \z_j,
\label{eq:tangentCondition}
\end{equation}
where $f_j$ has polynomial dependence on internal and external momenta.
\subsection{Relation to unitarity cut surfaces}
As we will see, the condition Eq.\ \eqref{eq:tangentCondition} has a nice geometric interpretation in terms of unitarity cut surfaces, similar to what was observed \cite{Ita:2015tya} in unitarity-compatible IBP reduction.

Consider a $L$-loop Feynman diagram topology with $N$ internal propagators. For a subset $\Delta$ of all inverse propagators ${1,2,\dots,N}$, the \emph{unitarity cut surface} labeled by $\Delta$ is defined as the hypersurface of all points $(l_1, l_2, \dots , l_L)$ in the \emph{complex} loop momentum space which solves the \emph{generalized unitarity cut condition},
\begin{equation}
z_i=0, \quad \forall i \in \Delta \, . \label{eq:unitarityCutCondition}
\end{equation}

Notice that generalized unitarity cuts differ from traditional unitarity cuts in QFT textbooks, treated by e.g.\ the optical theorem, Cutkosky rules \cite{Cutkosky:1960sp}, and the largest time equation \cite{Veltman:1963th,Remiddi:1981hn}, in several respects,
\begin{enumerate}
  \item Complex rather than real loop momentum space is considered. In particular, the energy component of the cut loop momentum is not required to be (real) positive. The algebraic closure of the complex field guarantees that the unitarity cut condition has solutions for generic external momenta, as long as not too many propagators are cut (which will be assumed to be the case for the rest of the paper).\footnote{For example, for one-loop topologies in $4$ spacetime dimensions, the unitarity cut condition can always be solved when $4$ or fewer propagators are cut.}
  \item There is no direct connection with the discontinuity of Feynman diagrams across branch cuts.
\end{enumerate}

When $\Delta$ contains all inverse propagators, i.e. when the unitarity cut condition sets all the propagators on-shell, the unitarity cut is called a maximal cut. Many problems, such as integrand construction and integration by parts, are simplest at the level of the maximal cut. When a proper subset of propagators are set on-shell, we have a non-maximal cut.

In general, for a diagram topology with $N$ propagators, there are $2^N$ different unitarity cut surfaces, since we may choose each propagator to be either cut or not cut. Since this paper is concerned with differential equations w.r.t.\ external momenta, we define ``extended unitarity cut surfaces'' which is embedded in the space of not only loop momenta but also external momenta. Here the loop momenta are still allowed to be complex but external momenta are required to be real. As usual, the surface is defined by the cut condition, Eq.\ \eqref{eq:unitarityCutCondition}.

For any of the $2^N$ extended unitarity cut surfaces $\Delta$ with $z_k \in \Delta$, the cut condition sets $z_k=0$, while the condition for the absence of doubled propagators, Eq.\ \eqref{eq:tangentCondition}, becomes
\begin{equation}
\left( \beta^\mu_i \frac{\partial}{\partial p_i^\mu} + v^\mu \frac{\partial}{\partial l^\mu}  \right) \z_k = f_k\, \z_k = 0 \, .
\label{eq:tangentCond}
\end{equation}
Therefore, the expression
\begin{equation}
\beta^\mu_i \frac{\partial}{\partial p_i^\mu} + v^\mu \frac{\partial}{\partial l^\mu},
\end{equation}
which we refer to as a ``DE vector'', is a tangent vector to every extended unitarity cut surface embedded in the space of both internal and external momenta. The loop part,
\begin{equation}
v^\mu \frac{\partial}{\partial l^\mu},
\label{eq:ibpVec}
\end{equation}
is called an ``IBP vector''. In the special case that the DE vector has no external momentum derivative, the IBP vector itself is a tangent vector to unitarity cut surfaces, and is used in unitarity-compatible IBP reduction. Computational algebraic geometry is used to find IBP vectors in the literature, and will also be used in this paper to find DE vectors.

\section{Detailed formalism in inverse propagator coordinates}
\label{sec:inverse}
\subsection{Inverse propagator coordinates}
\begin{figure}
\centering
 \includegraphics[width=0.9\textwidth]{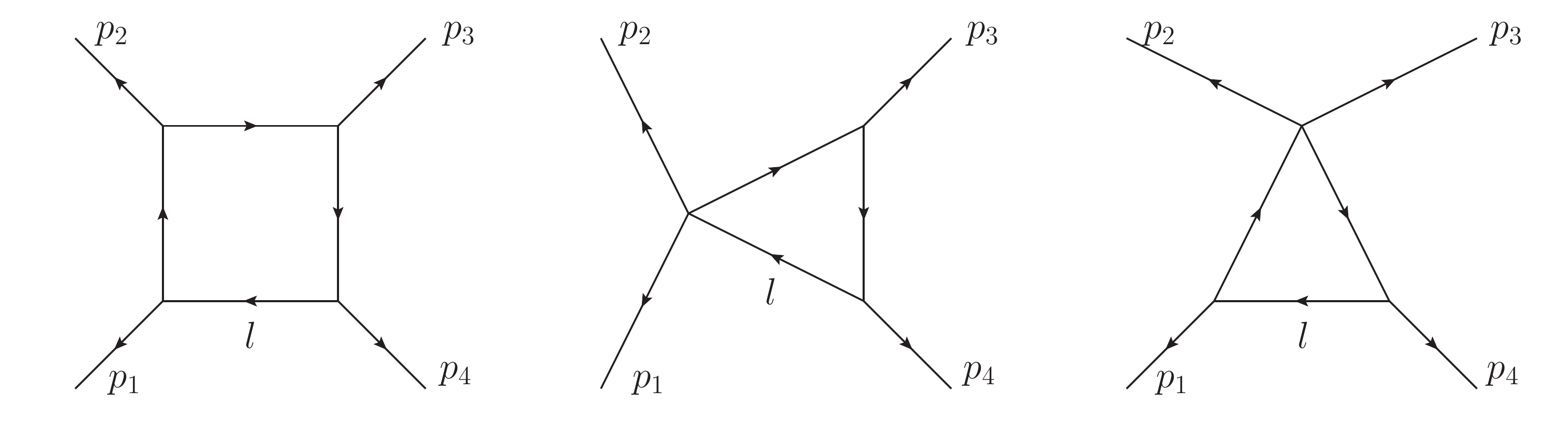}
\caption{Left: the scalar box integral with massless lines, $\mathcal I_{\rm box}$. Center: the $s$-channel scalar triangle integral, $\mathcal I_{\rm triangle}^{(s)}$. Right: the $t$-channel scalar triangle integral, $\mathcal I_{\rm triangle}^{(t)}$.}
\label{fig:boxMasters}
\end{figure}

We re-examine the simple example of the one-loop box \cite{Bern:1993kr} using our method. We assume that all internal and external lines are massless. The scalar box integral with some numerator $\mathcal N$, shown in the leftmost diagram of Fig.\ \ref{fig:boxMasters}, is
\begin{equation}
\mathcal I_{\rm box} = \int d^d l \, \frac{\mathcal N}{\prod_{j=1}^4 \z_j},
\end{equation}
with
\begin{align}
\z_1 &= l^2, \label{eq:z1def}\\
\z_2 &= (l-p_1)^2, \label{eq:z2def} \\
\z_3 &= (l-p_1-p_2)^2, \label{eq:z3def} \\
\z_4 &= (l-p_1-p_2-p_3)^2 = (l+p_4)^2 \, .  \label{eq:z4def}
\end{align}
The kinematic invariants are
\begin{align}
(p_1 + p_2)^2 &= s, \\
(p_2 + p_3)^2 &= t, \\
(p_1 + p_3)^2 &= u = -(s+t) \, .
\end{align}
It is well known that after IBP reduction, there are $3$ master integrals. For our purposes, they are conveniently chosen as the scalar box $\mathcal I_{\rm box}$, the $s$-channel triangle $\mathcal I_{\rm triangle}^{(s)}$, and the $t$-channel triangle integral $\mathcal I_{\rm triangle}^{(t)}$, shown in Fig.\ \ref{fig:boxMasters}.\footnote{Although there are $4$ possible daughter triangles of the box, the two $s$-channel triangles are identical after integration, and the same is true for the two $t$-channel triangles.}

The loop integral can be parameterized in the inverse propagator coordinates, in either $4$ or $d$ dimensions. This parameterization goes back to Cutkosky's proof of the cutting rules \cite{Cutkosky:1960sp}, and has been systematically studied by Baikov for the $d$-dimensional case \cite{Baikov:1996rk, Baikov:1996iu, Grozin:2011mt}. More recently, this parameterization was used in \cite{Ita:2015tya, Larsen:2015ped} for unitarity-compatible IBP reduction. A detailed explanation of the Baikov representation recently appeared in \cite{Frellesvig:2017aai}, where a public code was made available, and applications to differential equations were discussed. We now derive this parameterization for the one-loop box, and refer the readers to the literature for the multi-loop case.

Using the Van Neerven-Vermaseren basis \cite{vanNeerven:1983vr}, the metric tensor $\eta^{\mu \nu}$ is written as the sum of a ``physical'' component in the 3-dimensional space spanned by external momenta, and a ``transverse'' component $\hat \eta^{\mu \nu}$ in the remaining $(d-3)$-dimensional space which is orthogonal to every external momentum,
\begin{equation}
\eta^{\mu \nu} = \sum_{i=1,2,3} \, \sum_{j=1,2,3} {(G^{-1})}_{ij} \, p_i^\mu p_j^\nu  + \hat \eta^{\mu \nu} ,
\label{eq:VNbasis}
\end{equation}
where we used the inverse of the Gram matrix $G$,
\begin{equation}
G_{ij} = p_i \cdot p_j = \frac 1 2
 \begin{pmatrix}
 \setlength{\arraycolsep}{48pt}
 0 & \ s & \ -(s+t) \\
 s & \ 0 & \ t \\
 -(s+t) & \ t & \ 0
 \end{pmatrix} \, .
\end{equation}
We will need the relations,
\begin{align}
l \cdot p_1 &= \frac 1 2 (\z_1 - \z_2) = \frac 1 2 \z_{12}, \label{eq:lp1}\\
l \cdot p_2 &= \frac 1 2 (\z_2 - \z_3 + s) = \frac 1 2 (\z_{23} + s ), \label{eq:lp2}\\
l \cdot p_3 &= \frac 1 2 (\z_3 - \z_4 - s) = \frac 1 2 (\z_{34} - s )\, \label{eq:lp3} .
\end{align}
In the above equations we have defined
\begin{equation}
\z_{ij} = \z_i - \z_j \, .
\end{equation}
From Eq.\ \eqref{eq:z1def},
\begin{equation}
\z_1 = \eta^{\mu \nu} l_\mu l_\nu = {(G^{-1})}_{ij} (l \cdot p_i) (l \cdot p_j) - \hat \mu^2 ,
\end{equation}
where $\hat \mu^2$ denotes the Euclidean norm of the $(d-3)$-dimensional transverse components of $l$,
\begin{equation}
\hat \mu^2 = - \hat \eta^{\mu \nu} l_\mu l_\nu \, .
\end{equation}
Substituting Eqs.\ \eqref{eq:lp1}-\eqref{eq:lp3} into the above equation, $\hat \mu^2$ is expressed in terms of the inverse propagators $\z_i$,
\begin{equation}
\hat \mu^2 = F(\z) = {(G^{-1})}_{ij} (l \cdot p_i) (l \cdot p_j) - \z_1,
\end{equation}
where $F$ is called the Baikov polynomial. For the one-loop box, it evaluates to
\begin{align}
F &= \frac{s t}{4(s+t)} - \frac{1}{2(s+t)} \big [ s(\z_1 + \z_3) + t(\z_2 + \z_4) \big] + \frac{1}{4st(s+t)} \big [ s^2 (\z_2 - \z_4 )^2 + t^2 ( \z_1 - \z_3 )^2 \nonumber \\
&\quad + 2 st (-\z_1 \z_2 - \z_2 \z_3 - \z_3 \z_4 -\z_4 \z_1 + 2 \z_1 \z_3 + 2 \z_2 \z_4) \big ] \, .
\end{align} 
The loop integration measure, multiplied by the propagators, is re-written as
\begin{align}
\Omega = \frac{d^d l}{\prod_j \z_j} &= \frac{d(l\cdot p_1) d(l\cdot p_2) d(l\cdot p_3)}{\sqrt{\det G} \cdot \prod_j \z_j} \, d^{d-3} \hat \mu \nonumber \\
&= \frac{d \z_{12} \, d \z_{23} \, d \z_{34}}{2^3 \sqrt{\det G} \cdot \prod_j \z_j} \cdot \frac{\pi^{(d-3)/2}}{\Gamma((d-3)/2)} {(\hat \mu^2)}^{(d-5)/2} d \hat\mu^2 \nonumber \\
&= \frac{\pi^{(d-3)/2}}{2^3 \Gamma((d-3)/2) \sqrt{\det G} \cdot \prod_j \z_j} d \z_{12} \, d \z_{23} \, d \z_{34} \, d \hat \mu^2 \, (\hat \mu^2)^{(d-5)/2} \nonumber \\
&\quad \times d\z_1 \, \delta \left( \z_1 - {(G^{-1})}_{ij} (l \cdot p_i) (l \cdot p_j) - \hat \mu^2 \right) \nonumber \\
&= \frac{\pi^{(d-3)/2}}{2^3 \Gamma((d-3)/2) \sqrt{\det G} \cdot \prod_j \z_j} d \z_{12} \, d \z_{23} \, d \z_{34} \, d \z_1 \, {F(\z)}^{(d-5)/2} \nonumber \\
&= \frac{\pi^{(d-3)/2}}{2^3 \Gamma((d-3)/2) \sqrt{\det G}} \frac{d\z_1}{\z_1}\, \frac{d\z_2}{\z_2}\, \frac{d\z_3}{\z_3}\, \frac{d\z_4}{z_4}\, {F(\z)}^{(d-5)/2} \, .
\label{eq:baikovBox}
\end{align}
In the second-to-last line of the above equations, we integrated out $\hat \mu^2$ against the delta function, and in the last line we made a linear transformation of integration variables with unit Jacobian. Eq.\ \eqref{eq:baikovBox} accomplishes the transformation of the loop momentum from the Lorentzian / Euclidean component coordinates to the inverse propagator coordinates.

We will adopt the differential form notation for IBP relations \cite{Ita:2015tya, Larsen:2015ped}, which helps to transform between different coordinate systems for the loop momenta. We write
\begin{equation}
\int d^d l \frac{1} {\prod_j \z_j} = \int \Omega,
\end{equation}
where $\Omega$ is a maximal differential form in the loop momentum space. For convenience, we define $\Omega$ to include both the integration measure and the propagators. The total divergence integral of any IBP vector, $v = v^\mu \partial_\mu$, is
\begin{equation}
\int d^d l \frac{\partial}{\partial l^\mu} \frac{v^\mu} {\prod_j \z_j} = \int d \left( v \circ \Omega \right) \, .
\label{eq:ibpCoordFree}
\end{equation}
\subsection{Vectors for differential equations}
We first discuss IBP vectors, i.e.\ the loop component of DE vectors. The IBP vector Eq.\ \eqref{eq:ibpVec} can be transformed into the new parameterization,
\begin{equation}
\sum_{\mu} v^\mu \partial_\mu = \sum_i v_i \frac{\partial}{\partial \z_i} \, .
\end{equation}
Since $v^\mu$ is required to have polynomial dependence on external and internal momenta, it is easy to show that in the new coordinates, $v_i$ must have polynomial dependence on the inverse propagators $\z$. Furthermore, $v^\mu \partial_\mu$ is required to satisfy rotational invariance in the $(d-3)$-dimensional transverse space. An example of such an expression is
\begin{equation}
v^\mu \partial_\mu =\hat \eta^{\mu \nu} l_\nu \partial_\mu ,
\end{equation}
with $\eta^{\mu \nu}$ defined in Eq.\ \eqref{eq:VNbasis}.
This vector, and in fact all rotational invariant vectors satisfy
\begin{equation}
v^\mu \partial_\mu \, \hat \mu^2 \propto \hat \mu^2,
\end{equation}
where the proportionality constant has polynomial dependence on $\z_i
$. In the inverse propagator coordinates, this means
\begin{equation}
v_i (\z) \frac{\partial}{\partial \z_i} F(\z) = v_F (\z) \cdot F(\z),
\label{eq:Fcondition}
\end{equation}
for some $v_F$ which is a polynomial in the $\z$ variable.\footnote{We thank Harald Ita for giving a two-loop version of this argument in private communications.} The above equation is a crucial criterion for a valid IBP vector, derived in a different way in \cite{Larsen:2015ped}.

The IBP relation from an IBP vector $v=v_i \partial_{\z_i}$ is obtained by substituting the last line of Eq.\ \eqref{eq:baikovBox} into the exterior derivative expression Eq.\ \eqref{eq:ibpCoordFree},
\begin{align}
\int d^d l \frac{\partial}{\partial l^\mu} \frac{v^\mu} {\prod_j \z_j}
&\propto \int \frac{d\z_0}{\z_0}\, \frac{d\z_1}{\z_1}\, \frac{d\z_2}{\z_2}\, \frac{d\z_3}{z_3}\, {F(\z)}^{(d-5)/2} \nonumber \\
&\quad \times \left( \frac{d-5}{2} v_F + \sum_j \left(\frac{\partial v_j}{\partial \z_j} - \frac{v_j}{\z_j} \right) \right) ,
\label{eq:ibpDivBaikov}
\end{align}
with $v_F$ defined in Eq.\ \eqref{eq:Fcondition}.

Setting the tensor numerator $\mathcal N$ to $1$ in Eq.\ \eqref{eq:extPlusInt}, the 2nd term in the square bracket is given by Eq.\ \eqref{eq:ibpDivBaikov}, while the first term is simply, suppressing the $i$ index,
\begin{align}
&\quad \int d^d l\, \beta^\mu \frac{\partial}{\partial p^\mu} \frac{1}{\prod_j \z_j} = \int d^d l \, \frac{1}{\prod_j \z_j} (-1) \sum_j \frac 1 {\z_j} \beta^\mu \frac{\partial \z_j}{\partial p^\mu} \nonumber \\
&\propto (-1) \int \frac{d\z_1}{\z_1}\, \frac{d\z_2}{\z_2}\, \frac{d\z_3}{\z_3}\, \frac{d\z_4}{\z_4}\, {F(\z)}^{(d-5)/2} \, \sum_j \frac 1 {\z_j} \beta^\mu \frac{\partial \z_j}{\partial p^\mu} \, .
\label{eq:extBaikov}
\end{align}
Adding \eqref{eq:extBaikov} and the vanishing integral Eq.\ \eqref{eq:ibpDivBaikov}, we obtain,
\begin{align}
&\quad \int d^d l\, \left[ \beta^\mu \frac{\partial}{\partial p^\mu} \frac{1}{\prod_j \z_j}  + \frac{\partial}{\partial l^\mu} \frac{v^\mu} {\prod_j \z_j} \right] \\
&= \int \frac{d\z_1}{\z_1}\, \frac{d\z_2}{\z_2}\, \frac{d\z_3}{\z_3}\, \frac{d\z_4}{z_4}\, {F(\z)}^{(d-5)/2} \left[ \frac{d-5} 2 v_F +\sum_j \frac{\partial v_j} {\partial \z_j} - \sum_j \frac 1 {\z_j} \left( v_j + \beta^\mu \frac{\partial \z_j}{\partial p^\mu}  \right) \right] \, .
\label{eq:finalDE}
\end{align}
The above expression has no doubled propagators when
\begin{align}
& v_k + \beta^\mu \frac{\partial \z_k}{\partial p^\mu} = f_k \, \z_k \\
\implies & v_k =  f_k \, \z_k - \beta^\mu \frac{\partial \z_k}{\partial p^\mu} ,
\label{eq:tangentCondZ}
\end{align}
for some polynomial $f_k=f_k(\z)$ for every inverse propagator $\z_k$. This is equivalent to Eq.\ \eqref{eq:tangentCond}.

In summary, in inverse propagator coordinates, a unitarity-compatible DE vector that does not lead to doubled propagators is required to satisfy two requirements, Eq.\ \eqref{eq:Fcondition} and \eqref{eq:tangentCondZ}. Combining these two equations gives
\begin{align}
\sum_j \left( f_j \, \z_j - \beta^\mu \frac{\partial \z_j}{\partial p^\mu} \right) \frac{\partial F(\z)}{\partial \z_j} &= v_F \, F(z) \\
\implies \sum_j f_j \, \z_j \frac{\partial F(\z)}{\partial \z_j} - v_F F(z) &= \beta^\mu \frac{\partial \z_j}{\partial p^\mu} \frac{\partial F(\z)}{\partial \z_j} \, .
\label{eq:syzygy}
\end{align}
Given the derivative against external momenta, $\beta^\mu \partial_\mu$, Eq.\ \eqref{eq:syzygy} is a polynomial equation in the unknown polynomials $f_j$ and $v_F$. This is an inhomogeneous version of the ``syzygy equation'' of Ref.\ \cite{Larsen:2015ped}.\footnote{We note that Ref.\ \cite{Frellesvig:2017aai} used a different, but related, polynomial equation to construct differential equations within the Baikov representation, while our hybrid approach involves objects such as $\partial \z_j / \partial p^\mu$, which are in turn expressed as polynomials in the $\z$ variables.} We use the computational algebraic geometry package {\tt SINGULAR} \cite{DGPS} to find a particular solution for $f_j$ and $v_F$ in Eq.\ \eqref{eq:syzygy}, which in turn fixes $v_k$ in Eq.\ \eqref{eq:tangentCondZ}. This allows us to compute the derivative of the loop integral w.r.t.\ external momenta using Eq.\ \eqref{eq:finalDE}.
\subsection{Evaluating DEs: one-loop box warm-up}
Consider the derivative with respect to $t = (p_2 + p_3)^2$,
\begin{equation}
\left.\frac{\partial}{\partial t}\right|_s = \beta^\mu \frac{\partial}{\partial p^\mu} = \left( \frac{1}{2(s+t)} p_1^\mu + \frac 1 {2t} p_2^\mu + \frac{s+2t}{2t(s+t)} p_3^\mu \right) \frac{\partial}{\partial p_3^\mu} ,
\label{eq:tDeriExplicit}
\end{equation}
which annihilates $s=(p_1+p_2)^2$ and keeps external momenta on-shell.
Combined with Eqs.\ \eqref{eq:z1def}-\eqref{eq:z4def}, we find the following expressions for the RHS of Eq.\ \eqref{eq:syzygy},
\begin{align}
\beta^\mu \frac{\partial \z_4}{\partial p^\mu} &= \frac{-t(\z_1 + \z_3 - 2 \z_4) + s(t - \z_2 + \z_4) }{2t(s+t)} , \\
\beta^\mu \frac{\partial \z_i}{\partial p^\mu} &= 0, \quad i=1,2,3 \, .
\end{align}

Solving Eq.\ \eqref{eq:syzygy} for the unknown polynomials $f_i$ and $v_F$, we find the following solution, which can be easily checked,
\begin{align}
f_1 = f_2 = f_3 &= \frac{-t(\z_1 - 2\z_2 + \z_3) + s(t + \z_2 - \z_4)}{2t^2 (s+t)}, \nonumber \\
f_4 &= \frac{t(2t-\z_1 + 2\z_2 -\z_3) + s(3t + \z_2 -\z_4)}{2t^2(s+t)}, \nonumber \\
v_F &= \frac{t(\z_1 - 2 \z_2 + \z_3) - s (t + \z_2 -\z_4)} {t^2 (s+t)} ,
\end{align}
which in turn fixes $v_k$ in Eq.\ \eqref{eq:tangentCondZ}. Substituting the results into Eq.\ \eqref{eq:finalDE} gives the $t$-derivative of the box integral in a form that has no doubled propagators,
\begin{align}
&\quad \left. \frac{\partial}{\partial t} \right|_s \mathcal I_{\rm box} = \left. \frac{\partial}{\partial t} \right|_s \left[ \int d^d l \frac{1}{\prod_j \z_j} \right] \nonumber \\
&= \int d^d l \frac{1}{\prod_j \z_j } \left[ - \frac{s(1+\epsilon)+t}{t(s+t)} - \frac{\epsilon (s+2t)\z_2}{t^2(s+t)} + \frac{\epsilon}{t(s+t)} \left( \z_1 + \z_3 + \frac s t \z_4 \right) \right ] \nonumber \\
&= - \frac{s(1+\epsilon)+t}{t(s+t)} \mathcal I_{\rm box} - \frac{2\epsilon}{t(s+t)} \mathcal I_{\rm triangle}^{(s)} + \frac{2\epsilon}{t(s+t)} \mathcal I_{\rm triangle}^{(t)},
\end{align}
where we set $d=4-2\epsilon$. The triangle integrals are one-scale integrals whose derivatives against $t$ can be fixed by simple dimensional analysis, so we omit the calculation. Transforming to the basis suggested in Refs.\ \cite{Henn:2014qga, Henn:2013pwa}, we obtain
\begin{equation}
\frac{\partial}{\partial t}
 \begin{pmatrix}
 st \, \mathcal I_{\rm box}  \\
 s \, \mathcal I_{\rm triangle}^{(s)}  \\
 t \, \mathcal I_{\rm triangle}^{(t)}
 \end{pmatrix}
=
 \begin{pmatrix}
 -\frac{s \epsilon}{t(s+t)} & \frac{-2\epsilon}{s+t} & \frac{2\epsilon s}{t(s+t)} \\
0 & 0 & 0  \\
0 & 0 & - \frac{\epsilon}{t}
 \end{pmatrix}
 \begin{pmatrix}
 st \, \mathcal I_{\rm box}  \\
 s \, \mathcal I_{\rm triangle}^{(s)}  \\
 t \, \mathcal I_{\rm triangle}^{(t)}
 \end{pmatrix},
\end{equation}
which agrees with the result obtained from the standard approach involving IBP reduction of integrals with doubled propagators. In this simple example, we directly obtain the DEs without any IBP reduction, but unitarity-compatible IBP reduction of tensor integrals is needed in more complicated cases.
\subsection{4D leading singularities}
Although this paper mainly concerns unitarity cuts in $d$ dimensions, the inverse propagator coordinates also make it easy to compute cuts in $4$ dimensions. Setting $d=4$ in Eq.\ \eqref{eq:baikovBox}, the maximal cut residue of the box integral $\mathcal I_{\rm box}$ is equal to
\begin{equation}
\frac{\pi^{1/2}}{2^3 \Gamma(1/2) \, \sqrt{\det G}} F(0)^{-1/2}
\propto \frac{1} {st} ,
\end{equation}
where $F(0) = F(\z_i = 0)$.
Therefore, $s t \, \mathcal I_{\rm box}$ is an integral whose leading singularity is a $\mathbb{Q}$ number with no dependence on $s$ or $t$. This is an important criterion for selecting candidate integrals with uniform transcendentality.

\begin{figure}
\centering
 \includegraphics[width=0.4\textwidth]{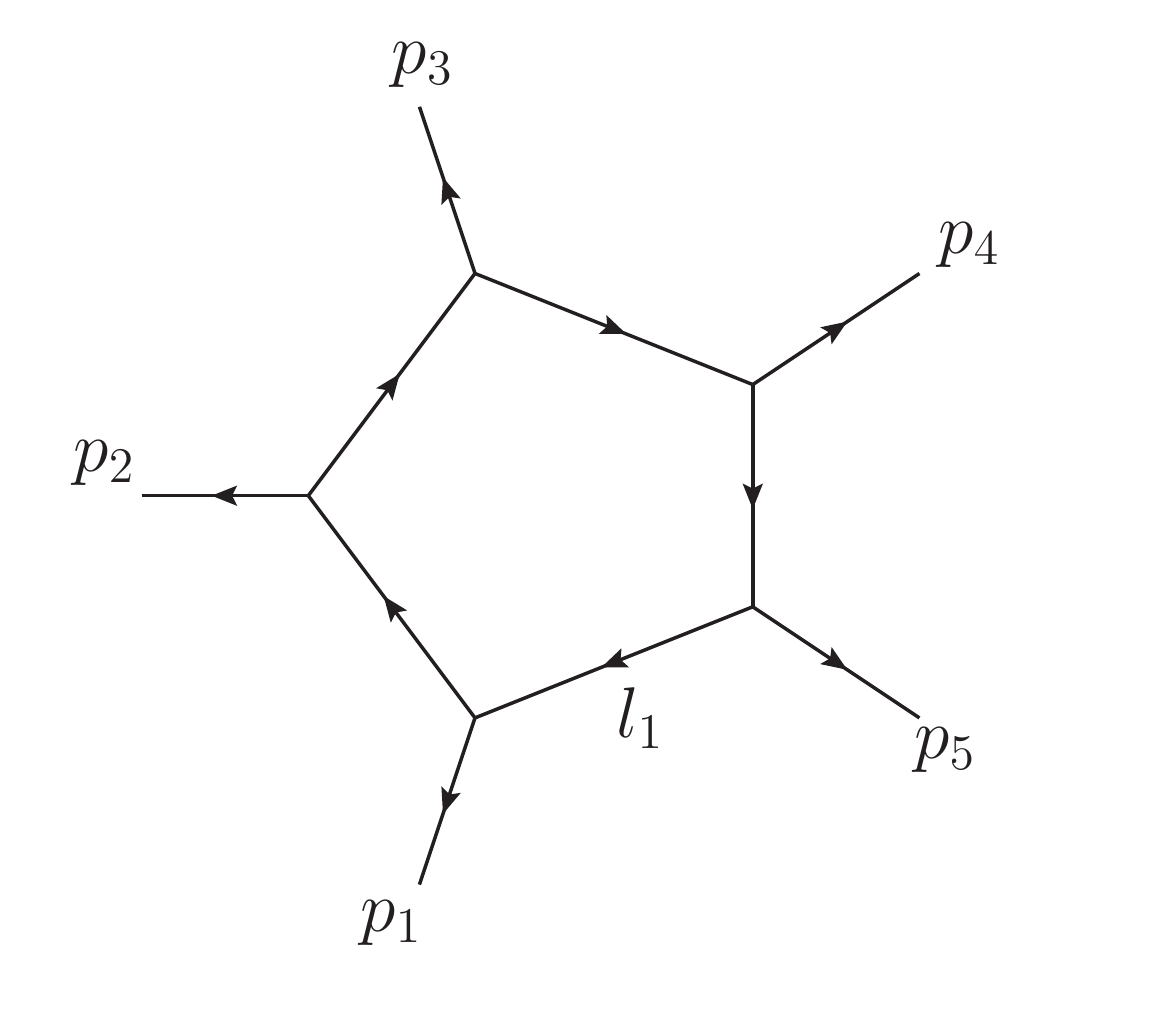}
\caption{The one-loop pentagon with massless lines.}
\label{fig:pentagon}
\end{figure}
We now translate the pentagon integrals with unit leading singularities \cite{ArkaniHamed:2010gh} to the inverse propagator coordinates, which serves as a building block for the master integral choice for the nonplanar pentabox. This loop topology is shown in Fig.\ \ref{fig:pentagon}. The five kinematic invariants can be chosen, in a cyclic invariant fashion, to be $s_{12}$, $s_{23}$, $s_{34}$, $s_{45}$, $s_{51}$, where $s_{ij} = (p_i + p_j)^2$. The Gram matrix is again defined as $G_{ij} = p_i \cdot p_j, 1 \leq i,j \leq 4$. In $D$ dimensions, the pentagon integral with a numerator $\mathcal N$ can be expressed in terms of the five inverse propagator coordinates, $\z_i = (l- \sum_{j=1}^{i-1} p_j)^2$,
\begin{equation}
\mathcal I_{\rm pentagon} = \frac{\pi^{(d-4)/2} \, \mathcal N}{2^4 \Gamma((d-4)/2) \sqrt{\det G}} \int \prod_{j=1}^5 \frac{d \z_j} {\z_j} F(\z)^{(d-6)/2},
\end{equation}
whereas in $4$ dimensions, the Baikov polynomial resides in a Dirac delta function since the $\z_i$'s become linearly dependent \cite{Ita:2015tya},
\begin{equation}
\mathcal I_{\rm pentagon}^{\rm 4d} = \frac{\pi^{(d-4)/2}\, \mathcal N}{2^4 \sqrt{\det G}} \int \prod_{j=1}^5 \frac{d \z_j} {\z_j} \delta \left( F(\z) \right) \, .
\label{eq:pentagon4D}
\end{equation}
The Baikov polynomial $F(\z)$ is invariant under a cyclic permutation of the indices, $\z_i \rightarrow \z_{i+1}, s_{ij} \rightarrow s_{i+1 \, j+1}$, modulo $5$.
Let us consider the $4$-dimensional cut $\z_1 = \z_2 = \z_3 = \z_4 = 0$, which localizes Eq.\ \eqref{eq:pentagon4D} to, omitting constant factors,
\begin{equation}
\frac{\mathcal N}{\sqrt{\det G}} \int \frac{d\z_5}{\z_5} \delta \left( \frac{F(0)}{\z_5^+ \z_5^-} (\z_5 - \z_5^+) (\z_5 - \z_5^-) \right),
\label{eq:localizedZ5Integral}
\end{equation}
where $F(0)$ is again a shorthand for
\begin{equation}
F(\z_i = 0) = \frac 1 {16 \det G} s_{12} \, s_{23}\, s_{34}\, s_{45}\, s_{51},
\end{equation}
and $\z_5^\pm$ are the two solutions for $\z_5$ on the cut,
\begin{equation}
z_5^\pm = \frac{1}{2 (s_{45} - s_{12} - s_{23})} \left( s_{12} s_{23} - s_{23} s_{34} + s_{34} s_{45} - s_{12}s_{51} + s_{45}s_{51} \mp 4 \sqrt{\det G} \right) \, .
\end{equation}
Choosing
\begin{align}
\mathcal N_{\z_1 = \z_2 =\z_3 =\z_4 =0} = \mathcal N^{\rm chiral} (\z_5) = \frac{F(0)}{\sqrt{\det G}} \cdot \frac{\z_5 - \z_5^+}{\z_5^+} ,
\label{eq:pentagonAnsatz}
\end{align}
then Eq.\ \eqref{eq:localizedZ5Integral} produces the leading singularity $1$ on the cut solution $\z_5 = \z_5^-$, and $0$ on the other cut solution $\z_5 = \z_5^+$. One important feature which makes Eq.\ \eqref{eq:pentagonAnsatz} a valid ansatz is that the $\z_5$-independent term, $-F(0) / \sqrt{\det G}$, is a cyclic invariant expression which is the same on all the five possible $4D$ maximal cuts. So Eq.\ \eqref{eq:pentagonAnsatz} can be promoted to the full uncut expression for $\mathcal N$ by combining the $\z$-dependent terms obtained on the individual cuts, and the result corresponds to the chiral pentagon integral of Ref.\ \cite{ArkaniHamed:2010gh} up to terms that vanish on $4$-dimensional box cuts.

\section{A non-planar five-point topology at two-loops}
\label{sec:pentabox}
\begin{figure}
\centering
 \includegraphics[width=0.5\textwidth]{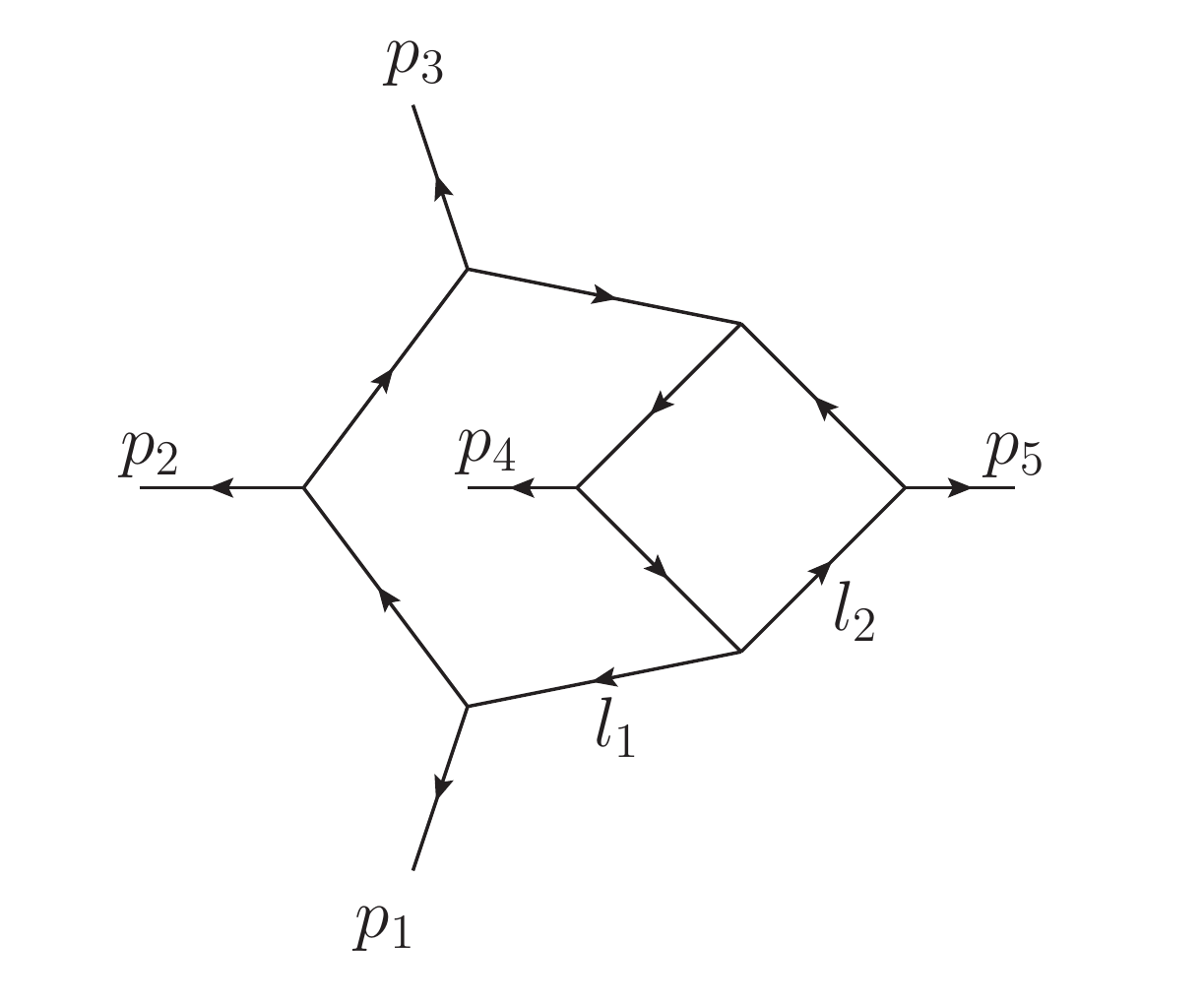}
\caption{The nonplanar pentabox integral with massless propagators and external legs with zero masses.}
\label{fig:pentaCrossBox}
\end{figure}

To illustrate our method for computing differential equations on unitarity cuts, we consider two-loop five-point scattering. While planar master integrals have been computed \cite{Gehrmann:2015bfy, Papadopoulos:2015jft}, the nonplanar counterpart remains unknown. Ref.\ \cite{Frellesvig:2017aai} suggested probing the properties of such integrals via the maximal cut. We focus on the nonplanar pentabox in Fig.\ \ref{fig:pentaCrossBox}, and compute differential equations at the maximal cut level.
The inverse propagators are
\begin{align}
\z_1 &= l_1^2, \quad \z_i = (l_1 - \sum_{j=1}^{i-1} p_j)^2, \ 2 \leq i \leq 4, \nonumber \\
\z_5 &= (l_1 + l_2 + p_4)^2, \quad \z_6 = (l_1 + l_2)^2, \nonumber \\
\z_7 &= l_2^2, \quad \z_8 = (l_2 - p_5)^2 ,
\end{align}
while the irreducible numerators are initially chosen as
\begin{equation}
\z_9 = (l_2 + p_1)^2, \quad \z_{10} = (l_2 + p_1 + p_2)^2, \quad \z_{11} = (l_2 - p_4 - p_5)^2 \, .
\label{eq:irredNums}
\end{equation}
Using inverse propagator coordinates, the loop integral can be written as, omitting overall factors,
\begin{equation}
\frac{1}{\det G} \int \prod_{j=1}^8 \frac{d\z_j}{\z_j} \prod_{j=9}^{11} d \z_j \, F(z)^{(d-7)/2} ,
\end{equation}
where $\det G$ is the Gram determinant for the external momenta, defined in the same way as for the one-loop pentagon, and $F(z)$ is defined by the $(d-4)$-dimensional components of $l_1$ and $l_2$, which are $\mu_1$ and $\mu_2$, by
\begin{equation}
F(z) = \mu_1^2 \mu_2^2 - ( \mu_1 \cdot \mu_2)^2 \, .
\end{equation}
The five independent kinematic invariants are defined in the same way as for the one-loop pentagon, as
\begin{equation}
s_{ij} = (p_i + p_j)^2, \quad (ij)=(12), (23), (34), (45), (51) \, .
\end{equation}
The loop integrals, up to an overall power of $s_{12}$, depend on the four dimensionless ratios,
\begin{equation}
\chi_{23} = \frac{s_{23}}{s_{12}}, \quad \chi_{34} = \frac{s_{34}}{s_{12}}, \quad \chi_{45} = \frac{s_{45}}{s_{12}}, \quad \chi_{51} = \frac{s_{51}}{s_{12}} \, .
\end{equation}
We perform IBP reduction at the maximal level for tensor numerators with up to $6$ total powers of $\z_9$, $\z_{10}$ and $\z_{11}$, using an in-house implementation of the algorithm of \cite{Larsen:2015ped}. This involves solving a syzygy equation, which is the same as our Eq.\ \eqref{eq:syzygy} with the RHS set to zero, to find vectors whose total divergences generate IBP relations without doubled propagators. Three master integrals, $\mathcal I_1$, $\mathcal I_2$, and $\mathcal I_3$, are found, with tensor numerators
\begin{equation}
 z_{11}^2, \quad \z_{11}, \quad 1,
\label{eq:pentaCrossBoxMasters}
\end{equation}
respectively.

Next, the derivatives w.r.t.\ the four scaleless kinematic invariants, $\partial / \partial \chi_{ij}$, are each written in terms of external momentum derivatives $\sim \beta^\mu (\partial / \partial p^\mu)$, in the fashion of Eq.\ \eqref{eq:tDeriExplicit}.
This fixes the RHS of Eq.\ \eqref{eq:syzygy}. On a unitarity cut, Eq.\ \eqref{eq:syzygy} is simplified, allowing a solution to be found by the computer more quickly.\footnote{If a particular $\z_k$ is set to zero, the $f_k$ term can simply be dropped from the LHS of the equation, while the RHS of the equation needs to be evaluated \emph{before} $\z_k$ (and any other inverse propagators in the cut) is set to zero, since these two operations do not commute.} Finding one particular solution for $f_j$ and $v_F$ using {\tt SINGULAR} gives us the needed ingredients to compute the derivative of the loop integral in Eq.\ \eqref{eq:diffPlusDiv}. We obtain a combination of tensor integrals without doubled propagators, which are then reduced to the three master integrals using the IBP reduction procedure described above. The end results are the differential equations relating the three master integrals with their partial derivatives against the four scaleless kinematic invariants,
\begin{equation}
\frac{\partial \mathcal I_A}{\partial \chi_{ij}} = M_{AB}^{ij} \, \mathcal I_B \, .
\label{eq:deMatrix}
\end{equation}
The full results for the matrix $M_{AB}^{ij}$ are attached in an ancillary file {\tt pentaCrossBox.m} in the {\it Mathematica} format. We will elaborate on the computation techniques in Subsection \ref{subsec:finiteField}. A sample component is
\begin{align}
M_{31}^{23} &= \frac {(1 + 4 \epsilon) \left(\chi _{23}-\chi _{45}+1\right)}
{\chi _{45} \left(-\chi _{23}+\chi _{45}+\chi _{51}\right)}
\left( \chi _{34} \chi _{23}-\chi _{23}-\chi _{34} \chi _{45}+2 \chi _{45}+\chi _{45} \chi _{51}+\chi _{51} \right) / \nonumber \\
& \quad \big [ \chi _{23}^2 \left(\chi _{34}-1\right){}^2+\left(\chi _{34} \chi _{45}-\left(\chi _{45}-1\right) \chi _{51}\right){}^2+2 \chi _{23} \nonumber \\
& \quad  \left(-\chi _{45} \chi _{34}^2+\chi _{45} \chi _{34}+\chi _{45} \chi _{51} \chi _{34}
+\chi _{51} \chi _{34}+\chi _{45} \chi _{51}-\chi _{51}\right) \big ] ,
\label{eq:sampleResult}
\end{align}
which have poles that can be identified with kinematic singularities, such as $\chi_{45} \propto (p_4 + p_5)^2$, and $(\chi_{23} - \chi_{45} + 1) \propto (p_1 + p_3)^2$. Since Eq.\ \eqref{eq:deMatrix} allows second derivatives to be computed, we are able to perform another check using the consistency condition \cite{Meyer:2016slj},
\begin{equation}
\frac{\partial}{\partial \chi_{ij}} \left( \frac{\partial \mathcal I_A}{\partial \chi_{kl}} \right) = \frac{\partial}{\partial \chi_{kl}} \left( \frac{\partial \mathcal I_A}{\partial \chi_{ij}} \right) \, .
\end{equation}
As is the case for the component shown in Eq.\ \eqref{eq:sampleResult}, the complete matrix is linear in $\epsilon$ but not proportional to $\epsilon$, because we have not yet transformed the result into a ``good" basis of master integrals with unit leading singularities. Such a ``good" basis can be found easily using the 4D cut method in \cite{Henn:2013pwa}. Roughly speaking, cutting the $l_2$ box sub-loop produces the Jacobian
\begin{equation}
\frac{1}{\z_5^\prime \z_5^{\prime \prime}}, \quad \text{where }  \z_5^\prime \equiv (l_1 + p_5)^2, \quad \z_5^{\prime \prime} \equiv (l_1 + p_4)^2 \, .
\end{equation}
So we can recycle the one-loop chiral pentagon expression Eq.\ \eqref{eq:pentagonAnsatz} to write down three tensor integrals, $\tilde {\mathcal I}_1$, $\tilde {\mathcal I}_2$, and $\tilde {\mathcal I}_3$ with unit leading singularities. Their numerators are (ignoring the dependence of the chiral pentagon numerator $\mathcal N^{\rm chiral}$ on $\z_1, \z_2, \z_3, \z_4$ which vanish on the maximal cut of the nonplanar pentabox),
\begin{align}
\mathcal N_1 &= \z_5^{\prime \prime} \, \mathcal N^{\rm chiral} (\z_5^\prime), \nonumber \\
\mathcal N_2 &= \z_5^{\prime} \, \mathcal N^{\rm chiral} (\z_5^{\prime\prime}), \nonumber \\
\mathcal N_3 &= s_{12} s_{23} \, \z_5^{\prime} \z_5^{\prime \prime}, \label{eq:5ptMasters}
\end{align}
respectively.
The first two of these integrals are among the nonplanar $\mathcal N=4$ SYM integrands for this particular topology given in \cite{Bern:2015ple}.

The new basis is related to the old one via a matrix $T$,
\begin{equation}
\tilde{\mathcal I}_A = T_{AB} \mathcal I_B,
\end{equation}
and the differential equations in the new basis,
\begin{equation}
\frac{\partial \tilde{\mathcal I}_A}{\partial \chi_{ij}} = \tilde M_{AB}^{ij} \, \tilde{\mathcal I}_B
\label{eq:deMatrixNew}
\end{equation}
are related to the old system, Eq.\ \eqref{eq:deMatrix}, by the transformation formula,
\begin{equation}
\tilde M^{ij} = T M^{ij} T^{-1} + \frac{\partial T}{\partial \chi_{ij}} T^{-1} \, .
\end{equation}
After the transformation, the matrices are proportional to $\epsilon = -(d-4)/2$, and now involve not only polynomials but also square roots of the Gram determinant. The square roots are eliminated by switching to momentum-twistor variables \cite{Hodges:2009hk} using the parameterization given in Appendix (A.2) of Ref.\ \cite{Badger:2013gxa}. The momentum-twistor variables, $x_i$ with $i=1,2,3,4,5$, are related to the usual kinematic invariants by
\begin{eqnarray}
x_1&=&s_{12}, \nonumber\\
x_2&=&\frac{s_{12} \left(s_{23}-s_{15}\right)+s_{23} s_{34}+s_{15} s_{45}-s_{34} s_{45}-\tr_5}{2
   s_{34}}, \nonumber\\
x_3&=&\frac{\left(s_{23}-s_{45}\right) \left(s_{23} s_{34}+s_{15} s_{45}-s_{34}
   s_{45}-\tr_5\right)+s_{12} \left(s_{15}-s_{23}\right) s_{23}+s_{12} \left(s_{15}+s_{23}\right)
   s_{45}}{2 \left(s_{12}+s_{23}-s_{45}\right) s_{45}}, \nonumber \\
x_4&=&-\frac{s_{12} \left(s_{23}-s_{15}\right)+s_{23} s_{34}+s_{15} s_{45}-s_{34}
   s_{45}+\tr_5}{2 s_{12} \left(s_{15}-s_{23}+s_{45}\right)}, \nonumber\\
x_5&=&\frac{\left(s_{23}-s_{45}\right) \left(s_{12}
   \left(s_{23}-s_{15}\right)+s_{23} s_{34}+s_{15} s_{45}-s_{34} s_{45}+\tr_5\right)}{2 s_{12} s_{23}
   \left(-s_{15}+s_{23}-s_{45}\right)},
\label{eq:twistor}
\end{eqnarray}
where $\tr_5$ is defined via the Gram determinant,
\begin{equation}
\tr_5 = 4 \sqrt{\det G} = 4 \sqrt{\det p_i \cdot p_j} \, .
\end{equation}
Since there are only $4$ dimensionless ratios of kinematic invariants, we will fix $x_1 = s_{12}=1$, effectively only looking at the dependence of the integrals on $x_2, x_3, x_4, x_5$.
After re-writing the differential equations in terms of the above momentum-twistor variables, we used the {\tt CANONICA} software package \cite{Meyer:2017joq} to transform the differential equation into a form that contains ``dlogs'' \cite{Henn:2013pwa} (a few seconds of computation time is used),
\begin{equation}
d \vec I = \epsilon \sum_{i=1}^{11} \mathbb M_i \vec I \, d \log s_i \, .
\end{equation}
In the above equation, $\vec I$ is a column vector consisting of the $3$ maximal-cut master integrals of unit leading singularities, $\tilde {\mathcal I}_1$, $\tilde {\mathcal I}_2$, and $\tilde {\mathcal I}_3$, defined via the tensor numerators in Eq.\ \eqref{eq:5ptMasters}. The $s_i$ variables are the so called symbol letters, with polynomial dependence on the momentum-twistor variables. The $3 \times 3$ matrices $\mathbb M_i$ are purely numerical, with no dependence on the dimension $d$ or the kinematic / momentum-twistor variables. The explicit expressions for the symbol letters are,
\begin{align}
s_1 &= x_2, \quad s_2 = x_3, \quad s_3 = x_2 + x_3, \nonumber \\
s_4 &= x_4, \quad s_5 = x_4 - x_5, \quad s_6 = -1 + x_5, \nonumber \\
s_7 &= -1 + x_4 + x_2 x_4 - x_2 x_5, \nonumber \\
s_8 &= 1 + x_2 x_5, \quad s_9 = -x_3 + x_2 x_5 + x_3 x_5, \nonumber \\
s_{10} &= -x_3 + x_2 x_4 + x_3 x_4 + x_2 x_3 x_4 - x_2 x_3 x_5, \nonumber \\
s_{11} &= -x_3 + x_2 x_4 + 2 x_3 x_4 + x_2 x_3 x_4 - x_2 x_3 x_5 - x_2 x_4 x_5 - x_3 x_4 x_5, \label{eq:5ptSymbols}
\end{align}
and the explicit expressions for the $3\times 3$ matrices are,
\begin{align}
\mathbb M_1 &= \begin{pmatrix}
 0 & 0 & 0 \\
 0 & -1 & -1 \\
 0 & -1 & -1
\end{pmatrix}, \quad
\mathbb M_2 = \frac 1 {16} \begin{pmatrix}
 -8 & -64 & 0 \\
 -3 & -24 & 0 \\
 1 & 8 & 0
\end{pmatrix}, \quad
\mathbb M_3 = \frac 1 {16} \begin{pmatrix}
 -8 & 0 &64 \\
 -1 & 0 & 8 \\
 3 & 0 & -24
\end{pmatrix}, \nonumber \\
\mathbb M_4 &= \begin{pmatrix}
 -1 & 0 & 0 \\
 0 & -1 & 0 \\
 0 & 0 & -1 
\end{pmatrix}, \quad
\mathbb M_5 = \frac 1 {4} \begin{pmatrix}
 0 & -16 & 16 \\
 -1 & -6 & -2 \\
 1 & -2 & -6
\end{pmatrix}, \quad
\mathbb M_6 = \frac 1 {16} \begin{pmatrix}
 -8 & 64 & 0 \\
 3 & -24 & 0 \\
 -1 & 8 & 0
\end{pmatrix}, \nonumber \\
\mathbb M_7 &= \frac 1 {16} \begin{pmatrix}
 -8 & 0 & -64 \\
 1 & 0 & 8 \\
 -3 & 0 & -24
\end{pmatrix}, \quad
\mathbb M_8 = \frac 1 {4} \begin{pmatrix}
 8 & -32 & 32 \\
 -1 & 2 & -2 \\
 1 & -2 & 2
\end{pmatrix}, \quad
\mathbb M_9 = \frac 1 {16} \begin{pmatrix}
 -8 & 0 & -64 \\
 1 & 0 & 8 \\
 -3 & 0 & -24
\end{pmatrix}, \nonumber \\
\mathbb M_{10} &= \frac 1 {16} \begin{pmatrix}
 -8 & 64 & 0 \\
 3 & -24 & 0 \\
 -1 & 8 & 0
\end{pmatrix}, \quad
\mathbb M_{11} = 2 \begin{pmatrix}
 0 & 0 & 0 \\
 0 & 1 & 0 \\
 0 & 0 & 1
\end{pmatrix} \, .
\end{align}
This form of the maximal-cut differential equations indicate that the solutions are multiple polylogarithms involving the $11$ symbol letters in Eq.\ \eqref{eq:5ptSymbols}, with uniform transcendentality in the $\epsilon$ expansion, as explained in Refs.\ \cite{Henn:2013pwa, Henn:2014qga}.

\section{Computation techniques and timing comparisons}
\label{sec:comp}
\subsection{Double box and comparison with {\tt FIRE5}}
\begin{figure}
\centering
 \includegraphics[width=0.6\textwidth]{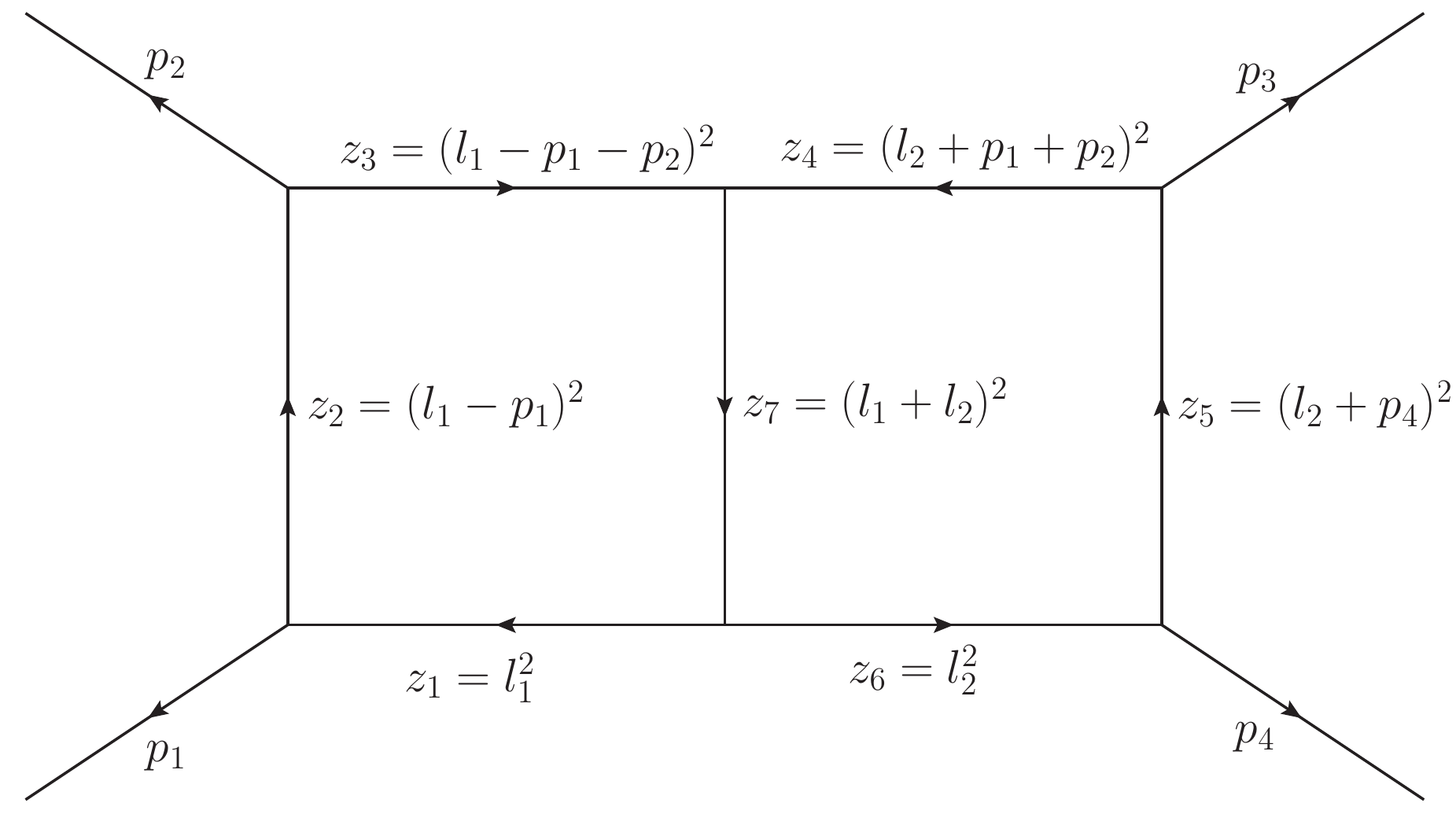}
\caption{The massless double box. The inverse propagators are labeled on the figure. The irreducible numerators are $z_8 = (l_1 + p_4)^2$ and $z_9 = (l_2 + p_1)^2$. The external kinematic invariants are $(p_1+p_2)^2=s$ and $(p_2+p_3)^2=t$.}
\label{fig:doubleBox}
\end{figure}
Having illustrated our method at the one-loop level in Section \ref{sec:inverse} and at the two-loop level in Section \ref{sec:pentabox}, we test the method for the double box and compare with existing methods. Being not too simple or too complicated, the double box topology can be handled by both traditional and unitarity-based methods, allowing for a meaningful comparison.

The double box with massless internal and external lines is shown in Fig.\ \ref{fig:doubleBox}, with kinematic variables defined in the caption. It is well known \cite{Smirnov:1999wz} that there are $2$ top-level master integrals, which may be chosen as the scalar integral $I_{\rm dbox}^s$ and the tensor integral $I_{\rm dbox}^t$,
\begin{equation}
I_{\rm dbox}^s = \int d^d l_1 d^d l_2 \, \frac{1}{\prod_{j=1}^7 z_j}, \quad
I_{\rm dbox}^t = \int d^d l_1 d^d l_2 \, \frac{z_9}{\prod_{j=1}^7 z_j} \, .
\end{equation}

With daughter topologies included, there are $12$ master integrals in total. Among them, $8$ master integrals are independent after accounting for the discrete symmetry,
\begin{align}
p_1 &\leftrightarrow p_4, \nonumber \\
p_2 &\leftrightarrow p_3, \nonumber \\
l_1 &\leftrightarrow l_2 \, . \label{eq:doubleBoxSymmetry}
\end{align}
which leaves the $7$ propagators invariant up to a permutation.
We re-compute the most non-trivial part of the system of differential equations, i.e.\ the $t$-derivative of the $2$ top-level master integrals, expressed in terms of the $8$ master integrals.

Our method is applied to a spanning set of $6$ different unitarity cuts,
\begin{align}
\Gamma_1 &= \{2,5,7\},\\
\Gamma_{2a} &= \{1,4,7\},\\
\Gamma_{2b} &= \{3,6,7\},\\
\Gamma_{3a} &= \{2,4,6,7\},\\
\Gamma_{3b} &= \{1,3,5,7\},\\
\Gamma_4 &= \{1,3,4,6\} \, .
\end{align}
The discrete symmetry Eq.\ \eqref{eq:doubleBoxSymmetry} is used to speed up the calculation by relating the cut $\Gamma_{2a}$ with $\Gamma_{2b}$, and $\Gamma_{3a}$ with $\Gamma_{3b}$. In particular, the IBP relations generated on one cut automatically become valid IBP relations on the related cut after the symmetry transformation. Not surprisingly, these are essentially the same cuts used in Ref.\ \cite{Larsen:2015ped} for IBP reduction of double box integrals.

By merging the results on the $6$ cuts, we reproduce the following differential equations,
\begin{align}
\frac{\partial}{\partial t} I_{\rm dbox}^s &= \frac {s(d-5)-t }{t(1+t)} I_{\rm dbox}^s + \frac{d-4}{t(1+t)} I_{\rm dbox}^t + \text{daughter topologies}, \nonumber \\
\frac{\partial}{\partial t} I_{\rm dbox}^t &= \frac{s(d-4)}{2(s+t)} I_{\rm dbox}^s - \frac{s(d-4)}{2t(1+t)} I_{\rm dbox}^t + \text{daughter topologies}, \label{eq:doubleBoxSampleResult}
\end{align}
where daughter topology integrals are \emph{fully computed} but omitted in the above sample results.

Both {\tt FIRE5} and our own unitarity-based code begin with ``preparation runs'' purely in {\it Mathematica} to process the diagram topology information supplied by the user. For the ``final runs'', {\tt FIRE5} is used in the C++ mode, while our own code is written in {\it Mathematica} and {\tt SINGULAR}. Only the final runs, which reflect the true computational complexities, are included in our timing comparison. The time required to obtain the results in Eq.\ \eqref{eq:doubleBoxSampleResult} is shown in Table \ref{tab:dboxTime}.
\begin{table}
  \centering
  \begin{tabular}{|c|c|c|}
    \hline
    Software & Time taken by final run \\ \hline
    {\tt FIRE5} & 141 seconds \\ \hline
    Own code & 37 seconds \\ \hline
  \end{tabular}
  \caption{Time required to obtain the results Eq.\ \eqref{eq:doubleBoxSampleResult} with full dependence on the $8$ master integrals including daughter topologies. The computation is performed on $1$ CPU core on a laptop computer with an Intel Core i5 processor (clock frequency 2.5 GHz). Finite field techniques and rational function reconstruction are not used for the timing comparison here, but will be used for the more complicated nonplanar five-point integrals.}
  \label{tab:dboxTime}
\end{table}

Though there could be more improvements by switching to statically compiled computer languages such as C++, our code already offers significant improvement in speed over a calculation based on {\tt FIRE5}, for the following possible reasons,
\begin{enumerate}
  \item The lack of doubled propagators reduces the number of different integrals that appear in IBP relations, because for combinatorial reasons, the number of integrals grow rapidly when propagators are allowed to be raised to higher powers.
  \item The use of unitarity cuts reduces the computational complexity.
  \item Besides solving syzygy equations, {\tt SINGULAR} is also used to solve the sparse linear system formed by the IBP relations. Unfortunately, due to lack of a controlled comparison, we do not know how this affects performance compared to {\tt Fermat} \cite{Smirnov:2008iw} used in the C++ version of {\tt FIRE5}.
\end{enumerate}

\subsection{Finite field techniques and rational function reconstruction}
\label{subsec:finiteField}
The computation of the differential equations for the nonplanar pentabox posed challenges in terms of CPU time and memory consumption. Given that the results, e.g.\ Eq.\ \eqref{eq:sampleResult}, are rational functions in $\chi_{ij}$, we use the rational function reconstruction technique of \cite{Peraro:2016wsq} to fit the analytic result from numerical inputs of $\chi_{ij}$.

The algorithm of \cite{Peraro:2016wsq} reconstructs multivariate rational functions in two steps, (i) fitting univariate rational functions, and (ii) fitting multivariate polynomials, using the input from many iterations of step (i). We use a simple private implementation which performs most of the work by exploiting the built-in capabilities of {\it Wolfram Mathematica}.\footnote{In {\it Mathematica 10}, step (i) is accomplished by the command {\tt FindSequenceFunction} with the option {\tt FunctionSpace -> "RationalFunction"}. Step (ii) is accomplished by the command {\tt InterpolatingPolynomial}. The latter command allows the option of computing in a finite field $\mathbb Z_p$.} For the nonplanar pentabox computation, step (i) requires $18$ kinematic points for each iteration, and $495$ iterations are performed to produce the input for step (ii). Step (ii) fits polynomials in $3$ variables, with degrees up to $8$. Finite field techniques are used to accelerate step (ii): using a large prime $p$, the full result can be constructed from its image in $\mathbb Z_p$ probabilistically using a minor modification of the extended Euclid algorithm \cite{Peraro:2016wsq, vonManteuffel:2014ixa}. After completing steps (i) and (ii), the fitted results are validated against new computations with additional random rational values of $\chi_{ij}$.

Here we give more information to quantify the performance gains from rational function reconstruction and finite field techniques. The computation of the differential equations for the nonplanar pentabox, with analytic dependence on the kinematic invariants, is very time consuming and does not finish after $48$ hours.\footnote{The computation gets stuck at the first stage, namely finding IBP-generating vectors that do not cause doubled propagators.} However, with (rational) numerical kinematic invariants, the computation finishes in a few seconds per kinematic point on a modern computer. A total of $28$ hours is used in evaluating differential equations on $8910$ kinematic points and reconstructing the full analytic results. The last step of the calculation, i.e. multivariate polynomial fitting, dramatically benefits from finite field techniques and takes about $52$ seconds to reconstruct all the $36$ entries of the four $3\times 3$ matrices. In contrast, when finite field techniques are turned off, about $395$ seconds are needed to reconstruct only \emph{one} of these $36$ entries (with the computation aborted afterwards), which is slower by more than $2$ orders of magnitude.

\section{Conclusions}
\label{sec:conclusions}
We have proposed a new method for constructing differential equations for Feynman integrals, which avoids generating integrals with doubled propagators, instead producing tensor integrals to be reduced by unitarity-compatible IBP reduction. In fact, for the simplest cases such as the one-loop box, no IBP reduction is needed at all. Our method allows constructing differential equations from a spanning set of unitarity cuts in $d$ dimensions, with IBP reduction also performed on the cuts.

Applying our method to the nonplanar pentabox, we obtained the homogeneous differential equations on the maximal cut in Henn's canonical form. This allows us to confirm that the master integrals, at least when evaluated on the maximal cut, are multiple polylogarithms with uniform transcendentality in the $\epsilon$ expansion. We have extracted the $11$ symbol letters, which are polynomials of momentum-twistor variables.

We also demonstrated that finite field techniques and rational function reconstruction, which are emerging as new tools in studying scattering amplitudes \cite{Peraro:2016wsq, vonManteuffel:2014ixa, vonManteuffel:2016xki}, are useful in computing differential equations for Feynman integrals.

There are several possible directions for follow-up studies. One direction is extending the calculation to other nonplanar five-point topologies, which can be done straightforwardly. It would be desirable to construct an automated implementation of our method, perhaps as an extension to unitarity-compatible IBP reduction software packages, such as Azurite \cite{Georgoudis:2016wff}, since many computation steps can be shared. Eventually, we would like to construct full DEs for nonplanar two-loop five-point integrals, which are relevant for NNLO QCD corrections for $2\rightarrow 3$ scattering processes at the LHC \cite{Gehrmann:2015bfy, Papadopoulos:2015jft}.

\section{Acknowledgment}
We thank Zvi Bern and Harald Ita for enlightening discussions and comments on the manuscript, and Yang Zhang for enlightening discussions and sharing ideas for efficient implementations of unitarity-compatible IBP reduction. The work of MZ is supported by the Department of Energy under Award Number DE-{S}C0009937.

\bibliographystyle{JHEP}
\bibliography{DE}
\end{document}